# Appearance and disappearance of superconductivity in SmFe$_{1-x}$Ni$_x$AsO (x = 0.0 to 1.0)

Anand Pal[1, 2*], S. S. Mehdi[2], Mushahid Husain[2] and V. P. S. Awana[1, #]

[1]Quantum Phenomenon and Applications (QPA) Division, National physical Laboratory (CSIR)

Dr. K.S. Krishnan Marg, New Delhi-110012, India

[2]Department of Physics, Jamia Millia Islamia, University, New Delhi-110025, India,

## Abstract

Bulk polycrystalline Ni-substituted SmFe$_{1-x}$Ni$_x$AsO (x = 0.0 to 1.0) samples are synthesized by solid state reaction route in an evacuated sealed quartz tube. The cell volume decreases with increase of Ni content in SmFe$_{1-x}$Ni$_x$AsO, thus indicating successful substitution of smaller ion Ni at Fe site. The resistivity measurements showed that the spin-density-wave (SDW) transition is suppressed drastically with Ni doping and subsequently superconductivity is achieved in a narrow range of x from 0.04 to 0.10 with maximum T$_c$ of 9K at x = 0.06. For higher content of Ni (x $\geq$ 0.10), the system becomes metallic and superconductivity is not observed down to 2K. The magneto-transport [R(T)H] measurements exhibited the upper critical field [H$_{c2}$(0)] of up to 300kOe. The flux flow activation energy (U/k$_B$) is estimated ~98.37K for 0.1T field. Magnetic susceptibility measurements also confirms bulk superconductivity for x = 0.04, 0.06 and 0.08 samples. The lower critical field (H$_{c1}$) is around 100Oe at 2K for x = 0.06 sample. Heat capacity C$_P$(T) measurements exhibited a hump like transition pertaining to SDW in Fe planes at around 150K and an AFM ordering of Sm spins below temperature of 5.4K for ordered Sm spins [T$_N$(Sm)]. Though, the SDW hump for Fe spins disappears for Ni doped samples, the T$_N$ (Sm) remains unaltered but with a reduced transition height, i.e., decreased entropy. In conclusion, complete phase diagram of SmFe$_{1-x}$Ni$_x$AsO (x = 0.0 to 1.0) is studied in terms of its structural, electrical, magnetic and thermal properties.

PACS: 74.10.+v, 74.25.Ha, 74.70.Ad

Key Words: SmFe$_{1-x}$Ni$_x$AsO; X-ray diffraction; Magnetization; and Heat capacity

*Corresponding Author: e-mail-sandhu.anand@gmail.com

Telephone No.: +91-11-45609357

#Dr. V.P.S. Awana, Scientist, NPL, New Delhi-12, India

e-mail-awana@mail.nplindia.ernet.in: Web page- www.freewebs.com/vpsawana/

## 1. Introduction

The discovery of superconductivity in layered iron-based LaFeAsO$_{1-x}$F$_x$ at 26 K [1] and successive increment of the same to above 56 K by replacing La ion with other rare earth elements such as Ce, Pr, Sm, Nd   [2-8] has been the biggest surprise to condensed matter community in recent years. Interestingly 56K is the highest superconducting transition temperature (T$_c$) after the famous high T$_c$ cuprates (HTSC) [9, 10]. The recent discovery provides an excellent opportunity to all the physicists working on theory of superconductivity, primarily because not only the mysterious cuprates (HTSC) but now the Fe based pnictide compounds also join the exclusive club of non BCS (Bardeen Cooper Schriefer) superconductors. Interestingly, the oxy-pnictide (REFeAsO, RE = rare earths) and cuprates have enough similarities; such as their structure is more or less layered and superconductivity resides in FeAs layers of the former and in CuO$_2$ planes of the  later and other building blocks of their unit cells work only as the charge reservoir redox layers. Certainly, the superconductivity in oxy-pnictides makes a debate of the unconventional superconductivity, similar to that as for HTSCs.

The parent compound REFeAsO is non-superconducting in its undoped pristine state shows an anomaly in resistivity versus temperature curves at approximately 140K [1-8].  This anomaly has been attributed to the collective effect of a crystallographic phase transition  at ~150 K, and an static antiferromagnetic long range ordering (SDW) of the Fe spins at a slightly lower temperature of ~140 K. The structural phase transition from the tetragonal *P4/nmm* to the orthorhombic *Cmma* space group happens at around 150 K. After doping of carriers, the spin density wave behaviour of compound shifts to lower temperature. The superconductivity appears after disappearance of SDW. The carriers for the superconductivity in REFeAsO compound are doped by different routes; (i) by substitution of F$^{1-}$on the O$^{2-}$ sites, (ii) by Inducing oxygen deficiency (iii) by partial substitution of the trivalent Rare-earth (RE) ions by bi-or tetravalent cationic species and (iv) partially substitution of trivalent 3d metal at the Fe Site [11–17]. Though the (i), (ii) and (iii) routes calls for the indirect injection of mobile carriers in superconducting FeAs layers by redox mechanism through doping in REO, the (iv) route is direct injection of carriers by doping in FeAs superconducting layer itself. Interestingly direct injection of carriers by doping in CuO$_2$ planes did never work in case of HTSc cuprates.  This feature makes a marked difference of Iron based superconductors from cuprate, in which any substitutions or direct insertion of carriers into the $CuO_2$ planes could not bring about superconductivity.

Here, we report synthesis, structural detail, electrical, magneto transport and specific heat of the Ni-substituted $SmFe_{1-x}Ni_xAsO$ (x = 0.0 to 1.0). The carriers in this case are directly injected in to the superconducting FeAs layer by doping Ni at Fe site. All the studied compounds are crystallized in a tetragonal structure with space group *P4/nmm* and single phase within XRD detection limit. The superconducting transition temperature dependence on Ni-doping(x) established a dome-like curve with highest $T_C$ at 9K for the x = 0.06. Superconductivity is not seen for higher Ni content ($x \geq 0.10$), which is most probably due to over doping of carriers. The findings of presently studied $SmFe_{1-x}Ni_xAsO$ (x = 0.0 to 1.0) system are compared with our reported results on $SmFe_{1-x}Co_xAsO$ (x = 0.0 to 1.0) [12, 16].

## 2. Experimental

All the studied polycrystalline $SmFe_{1-x}Ni_xAsO$ (x = 0.0 to 1.0) samples were prepared through single step solid-state reaction route via vacuum encapsulation technique [12, 13]. High purity (~99.9%) Sm, As, $Fe_2O_3$, Fe, and Ni in their stoichiometric amount are weighed, mixed and ground thoroughly using mortar and pestle under high purity Ar atmosphere in glove box. The Humidity and Oxygen content in the glove box is less than 1 ppm. The mixed powders were palletized and vacuum-sealed ($10^{-4}$ Torr) in a quartz tube. These sealed quartz ampoules were placed in box furnace and heat treated at $550^oC$ for 12 hours, $850^oC$ for 12 hours and then at $1150^oC$ for 33 hours in continuum with slow heating rate. Finally furnace is allowed to cool down to room temperature naturally.

The crystal structure was analyzed by the powder X-ray diffraction patterns at room temperature using Rigaku X-ray diffractometer with Cu $K_\alpha$ radiation. The resistivity measurements were carried out by a conventional four-probe method on a quantum design Physical Property Measurement System (PPMS). Heat capacity and magnetization measurements were also carried out on Quantum Design PPMS (Physical property measurement system) with fields up to 14 Tesla.

## 3. Results and Discussion

Figure 1 Shows the observed and Rietveld fitted X-ray diffraction (*XRD*) pattern of the representative samples of SmFe$_{1-x}$Ni$_x$AsO ($x$ = 0.0 to 1.0) compounds. The Rietveld analysis of the room temperature X-ray diffraction pattern confirmed that all the studies sample are crystallized in the tetragonal phase with space group P4/*nmm* in analogy of other ZrCuSiAs type structure. The Rietveld refinement was performed using the *FULLPROF SUITE* program. All the samples are apparently single phase with some weak impurity lines (observed in Ni-2% sample), which are marked with * (for Sm$_2$O$_3$) and # (for Fe$_2$O$_3$) in XRD pattern. The details of co-ordinate positions and lattice parameters along with the quality of fitting parameter are listed in the Table1 and Table 2; these values are broadly in agreement with earlier reports [14-17]. The lattice parameters and the unit cell volume as function of Ni doping ($x$) are plotted in fig. 1 (b), with $x$ ranging from 0.0 to 0.25. It is observed that with Ni doping *a* parameter increase slightly, while *c* lattice parameter shrinks remarkably. The reduction in the *c*-parameter is a result of the decrement of the distance between Sm-As with electron doping due to increased Coulomb attraction between adjacent layers [15]. The injection of charge carrier in FeAs layers lead to increase the distance between Fe and As atom and As-Fe-As block thickness. The FeAs$_4$ becomes more homogeneous and As-Fe-As angle approaching to perfect tetrahedral angle [15]. The cell volume is reduced consequently by the incorporation of Ni doping at the Fe site. The observed XRD patterns for SmFe$_{1-x}$Ni$_x$AsO ($x$ = 0.0 to 1.0) compositions indicate that Ni substitutes successfully with full solubility at the Fe site in SmFeAsO.

Figure 2(a) depicts the temperature versus resistivity behaviour for the representative sample of SmFe$_{1-x}$Ni$_x$AsO series. The pure undoped SmFeAsO itself not a superconductor, it shows an anomaly in resistivity at temperature near 140 K. This resistivity anomaly is due to the collective effect of spin density wave (SDW) instability and the structural phase transition from tetragonal to orthorhombic phase [2,12]. 1% and 2% doping of Ni at Fe site decrease the SDW transition temperature sharply from 140K to 88 and 56K respectively. Further doping of Ni completely suppress the SDW transition and introduced superconductivity for x = 0.04, 0.06 and 0.08 respectively at 7.5, 9 and 6K. The superconducting transition temperature dependence on Ni-doping(x) demonstrates a dome-like curve with highest T$_C$ at 9K for the optimal doping of x = 0.06, shown inset of Fig.2(a). The SmFe$_{1-x}$Ni$_x$AsO seemingly has the narrower superconducting window in compare with SmFe$_{1-x}$Co$_x$AsO [18]. The resistivity behaviour of SmFe$_{0.96}$Ni$_{0.04}$AsO is metallic above 130K and below which it is semiconducting like up to 7.5K where it exhibits superconducting transition. The resistivity of SmFe$_{0.94}$Ni$_{0.06}$AsO shows metallic behaviour below room temperature down to superconducting onset. With further doping of Ni (x$\geq$ 0.10) superconductivity disappears, which is primarily due to the over doping of carriers. Sample with 10% Ni doping shows a small dip in resistivity around 4K, which doesn't shows zero resistivity down to 3K. Further, it is observed that as the Nickel content increase (x$\geq$ 0.10), the metallic behaviour of the compound become more prominent, and superconductivity is not observed. The increasing metallic behaviour of SmFe$_{1-x}$Ni$_x$AsO compound is evidence for the increment of the carrier concentration with Ni doping.

In order to determine the upper critical field of the superconducting samples SmFe$_{0.96}$Ni$_{0.04}$AsO and SmFe$_{0.94}$Ni$_{0.06}$AsO resistivity $\rho(T)$-$H$ are measured under various applied magnetic fields up to 100kOe. The resistivity versus temperature under applied field along with the upper critical field is depicted in Fig. 2(b) and 2(c) for x=0.04 and 0.06 respectively. It is clear from the $\rho(T)H$ data that the superconducting transition temperature shifts to lower temperature as applied magnetic field increases. The width of the transition becomes broader with increasing magnetic field. From fig. 2 (b) and 2(C),  it is clear that the rate of decrease of transition temperature with applied magnetic field of the Ni-doped oxypnictide superconductor is around 1 Kelvin per Tesla {dT$_c$/dH ~ 1K/T} which is far less in compare to other high Tc superconductor like, YBCO {dT$_c$/dH ~ 4K/T} and MgB$_2$ {dT$_c$/dH ~ 2K/T}. The less value of dT$_c$/dH indicates toward a high value of upper critical field (H$_{c2}$) in these compounds [18]. The upper critical field [H$_{c2}$(T)] values at Zero temperature are calculated by the extrapolation method using Ginzburg-Landau (GL) theory.  The H$_{c2}$(T) is determined using different criterion of H$_{c2}$(T) = H at which $\rho$ =90% , 50% and 10% of $\rho_{N,}$ where $\rho_N$ is the normal state resistivity. The Ginzburg-Landau equation is:-

$$H_{c2}(T)=H_{c2}(0)*[(1-t^2)/(1+t^2)]$$

Where, t = T/T$_c$ is the reduced temperature and H$_{c2}$(0) is the upper critical field at temperature Zero . The Ginzburg-Landau (GL) equation, which not only determines the H$_{c2}$ value at zero Kelvin [H$_{c2}$(0)], but also determines the temperature dependence of critical field for the whole temperature range. The variation of H$_{c2}$(T) with temperature for 4% and 6% Ni doped sample are shown in the Fig.2 (d) and 2(e) respectively. It is clear from Fig.2 (d) and 2(e) that the H$_{c2}$(0) reaches above 300 KOe with $\rho$ =90% criteria.

The temperature derivative of resistivity for the superconducting samples SmFe$_{0.96}$Ni$_{0.04}$AsO and SmFe$_{0.94}$Ni$_{0.06}$AsO at various applied magnetic field are shown in Fig. 2(f) and 2(g). The temperature derivative of resistivity gives a narrow intense peak at T$_c$ in Zero applied fields, which indicate good percolation path of superconducting grain. The resistivity peak is broadened under applied fields. The broadening of the d$\rho$/dT peak increases with applied magnetic field.

The broadening of resistivity in superconductors under applied magnetic field is due to the thermally activated flux flow (TAFF) [19, 20.]. The resistance in broaden region is caused by the creep of vortices, which are thermally activated. The temperature dependence of resistivity in this region is given by Arrhenius equation [19],

$$\rho(T,B)=\rho_0 \exp[-U_0/k_B\, T]$$

Where, $\rho_0$ is the field independent pre-exponential factor (here normal state resistance at 12 K $\rho_{12}$ is taken as $\rho_0$), $k_B$ is the Boltzmann's constant and $U_0$ is TAFF activation energy, which can be obtained from the slope of the linear part of an Arrhenius plot in low resistivity region. We have plotted experimental data (Blue symbol) as ln($\rho/\rho_{12}$) vs. T$^{-1}$ in fig 3. The best fitted (red line) to the experimental data gives value of the activation energy ranging from U$_0$/k$_B$ = 98.37K to 8.43K in the magnetic field range of 0.1 T to 10T. The magnetic field dependence of activation energy is shown in inset of Fig.3. The activation energy shows weak dependence i.e. U$_0$/k$_B$ ~ H$^{-0.35}$ at low field but strongly decreases as U$_0$/k$_B$ ~ H$^{-0.84}$ for higher field range. These values are in good agreement with previous reports [21].

Figure 4(a) shows the temperature dependence of the DC magnetization for the SmFe$_{1-x}$Ni$_x$AsO (*x* = 0.04 to 0.1) samples. The measurements were carried out under a magnetic field of 10Oe in the zero field cooled (*ZFC*) and field cooled (*FC*) measuring conditions. It is clear from the Fig. 4(a) that all the samples show superconducting diamagnetic signal in both *FC* and *ZFC* measurement. The optimally doped SmFe$_{0.94}$Ni$_{0.06}$AsO sample gives the strongest diamagnetic signal with highest superconducting transition temperature at 9 K. This result is in good agreement with the earlier reports [14]. The diamagnetic transition of these samples confirms the bulk superconductivity in the present samples. The inset of Fig. 4(a) shows only the first quadrant of the *M(H)* loop at 2, 3 and 5K. As we increase the applied magnetic field, the magnetization goes in negative direction but around 90 Oe at 2K the *M(H)* curve invert. This field of inversion is known as the lower critical field ($H_{C1}$), at this value the magnetic field starts to penetrate the superconductor. For the higher values of temperature (3 and 5K) obviously the field of inversion is deceased. This shows that the lower critical field of the optimally doped sample is around 90Oe at 2K. The lower critical field is smaller for $x = 0.04$ and 0.08 doped sample. The variation of Magnetization under applied magnetic field, M ($H$) for the $T_c$ sample (x = 0.06) at various temperature 2, 5 & 10 K are shown in the Fig. 4(b). The opening of $M(H)$ loop gives the clear evidence of superconductivity.

The temperature dependence of heat capacity ($C_p$) for the SmFeAsO and the SmFe$_{0.94}$Ni$_{0.06}$AsO are shown in the main panel of fig. 5. The absolute values of $C_p$ are quite close for both samples. The absolute value of $C_p$ at 280 K 97.85 J/molK, which is very close to the Dulong Petit value, 3nR J/molK at high temperature, where n is the number of atom and R is the gas constant [12, 22-23]. In SmFeAsO a hump observed in Cp data around the same temperature at which a metallic step observed in resistivity measurement. This hump completely disappears in SmFe$_{0.94}$Ni$_{0.06}$AsO sample. The cusp like shape in SmFeAsO heat capacity around 140K is due to the spin density wave (SDW) character exhibited by the compound. It is also known that besides SDW the ground state non-superconducting REFeAsO also exhibit structural phase transition. Heat capacity decreasing with further decreases in temperature and another peak is observed at 4.5 K, shown in inset of Fig. 5. The sharp peak at 4.5 K is due to the antiferromagnetic ordering of Sm$^{3+}$ ions. The peak height depends on both doping level and applied magnetic field. It's clear from the inset of fig.5 that as doping level increases peak height decreases. This indicates that the change in entropy related to the ordering of Sm$^{3+}$ ions in Ni doped superconducting samples is less then the same for pure SmFeAsO. By the polynomial interpolation fitting of heat capacity data we are calculating change in entropy for each transition [22]. We are using the equation $a$T+bT$^3$ to estimation the background contribution in specific heat for each transition. For SDW contribution we fit this equation in the temperature range from 160K to 110K excluding the region from 140K to 120K. Using the fitted value of the coefficient $a$ and b, we calculated the background curve for the whole temperature range from 160K to 110K. To calculate change in specific heat ($\Delta C_P$), calculated data is subtracted by the experimental data. The change in entropy related to peak evaluated by the relation, $\Delta S = \int (\Delta C_P / T) dT$ where $\Delta C_P$ change in specific heat. For the SDW contribution the change in entropy is found to be $\Delta S = \sim 0.379(5)$ J/molK. Same interpolation scheme are applied on other transition. To determine the entropy change for $C_p$ transition at 4.5K, the background are estimated from the above equation fit in temperature range 20K to2K excluding temperature range 10K to 3K. We estimated the entropy change $\Delta S = \sim 4.622(8)$, $\sim 3.234$ (4), and $\sim 3.058$ (1) J/molK for the SmFeAsO, SmFe$_{0.96}$Ni$_{0.04}$AsO, and SmFe$_{0.94}$Ni$_{0.06}$AsO respectively. The decrease in $\Delta S$ with Ni content is possibly due to the changed Sm$^{3+}$ AFM fluctuations in doped samples. The antiferromagnetic ordering of Sm$^{3+}$ ions is less stabilized for the Ni doped samples in comparison to pure SmFeAsO.

Summarily, we have synthesized pure phase Ni doped polycrystalline SmFe$_{1-x}$Ni$_x$AsO (x = 0.0 to 1.0) samples. The successive Ni doping at iron site suppress the SDW character and introduce bulk superconductivity in a narrow doping window from x = 0.04 to 0.10 with maximum $T_c$ of 9K for the SmFe$_{0.94}$Ni$_{0.06}$AsO sample. The Superconductivity is not seen for higher Ni doping on Fe site.

Authors would like to thank Director NPL Prof. R.C. Budhani for his keen interest and encouragement for the study. Anand Pal would like to thank CSIR for granting senior research fellowship.

**Table 1.** Wyckoff position for $SmFe_{1-x}Ni_xAsO$ (Space group: *P4/nmm*)

| Atom | Site | X | y | z |
|------|------|------|------|------|
| Sm | 2c | 0.25 | 0.25 | 0.139(6) |
| Fe/Ni | 2b | 0.75 | 0.25 | 0.50 |
| As | 2c | 0.25 | 0.25 | 0.655(4) |
| O | 2a | 0.75 | 0.25 | 0.00 |

**Table 2.** Reitveld refined parameters for $SmFe_{1-x}Ni_xAsO$

| Ni-% | a (A) | c (A) | Volume ($A^3$) | Rwp | $X^2$ | $T_M(K)$ | $T_C(K)$ |
|------|-------|-------|----------------|-----|-------|----------|----------|
| 0 | 3.937(2) | 8.492(1) | 131.64 | 3.7 | 1.97 | 144 | - |
| 1 | 3.937(8) | 8.487(1) | 131.61 | 2.91 | 0.973 | 88 | - |
| 2 | 3.937(7) | 8.480(3) | 131.50 | 3.02 | 1.07 | 56 | - |
| 4 | 3.938(4) | 8.471(2) | 131.40 | 2.87 | 0.847 | - | 7.5 |
| 6 | 3.939(1) | 8.459(2) | 131.26 | 3.79 | 2.21 | - | 9 |
| 8 | 3.940(2) | 8.450(5) | 131.19 | 3.96 | 1.82 | - | 6 |
| 10 | 3.941(2) | 8.444(4) | 131.17 | 3.86 | 1.95 | - | <3 |
| 12 | 3.94264 | 8.433(1) | 131.09 | 4.48 | 2.11 | - | - |
| 15 | 3.943(3) | 8.427(4) | 131.05 | 3.93 | 2.21 | - | - |
| 25 | 3.952(3) | 8.387(6) | 131.02 | 3.93 | 2.02 | - | - |
| 100 | 4.026(6) | 8.018(7) | 130.01 | 5.26 | 2.68 | - | - |

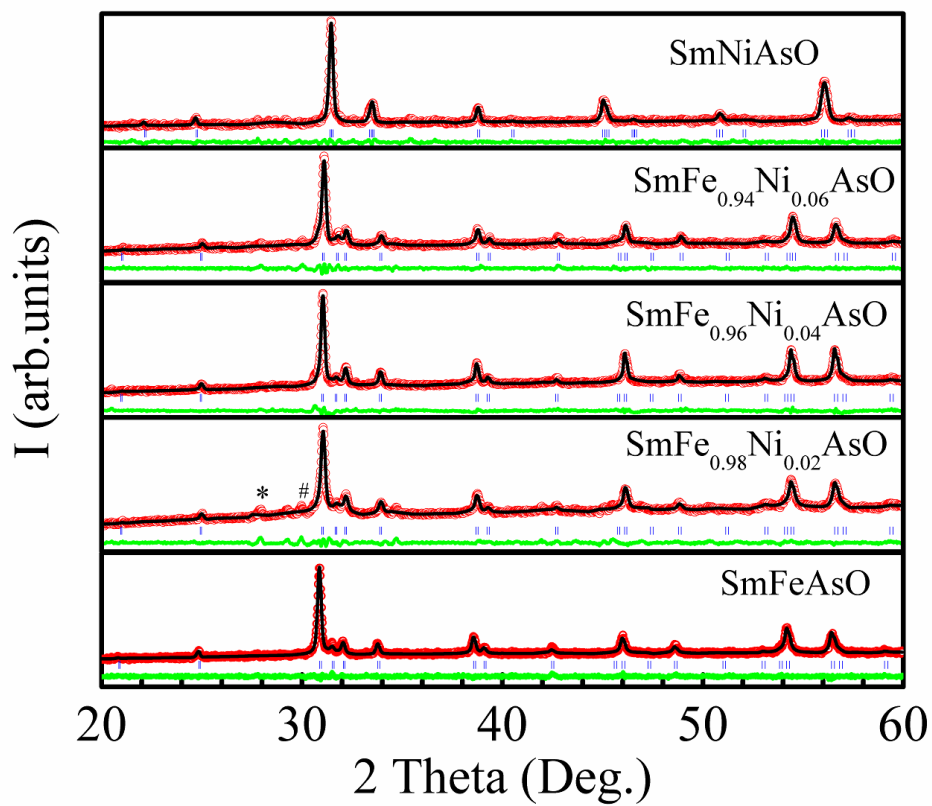

**Figure 1(a):** Observed and Rietveld fitted room temperature XRD patterns of representative samples of SmFe$_{1-x}$Ni$_x$AsO   ($x$ = 0.0, 0.02, 0.04, 0.06 & 1.0)

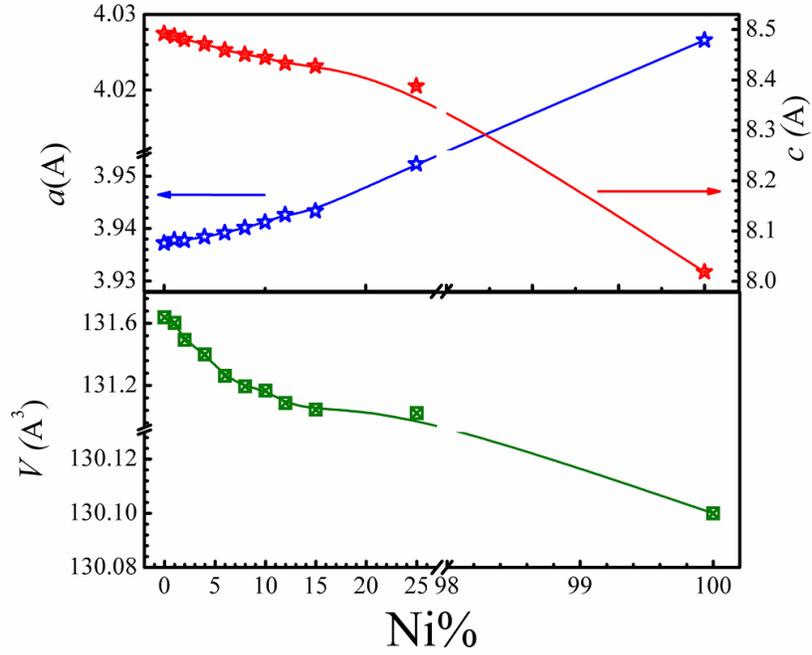

**Figure 1(b):** Variation of lattice parameter and unit cell volume with of Ni concentration.

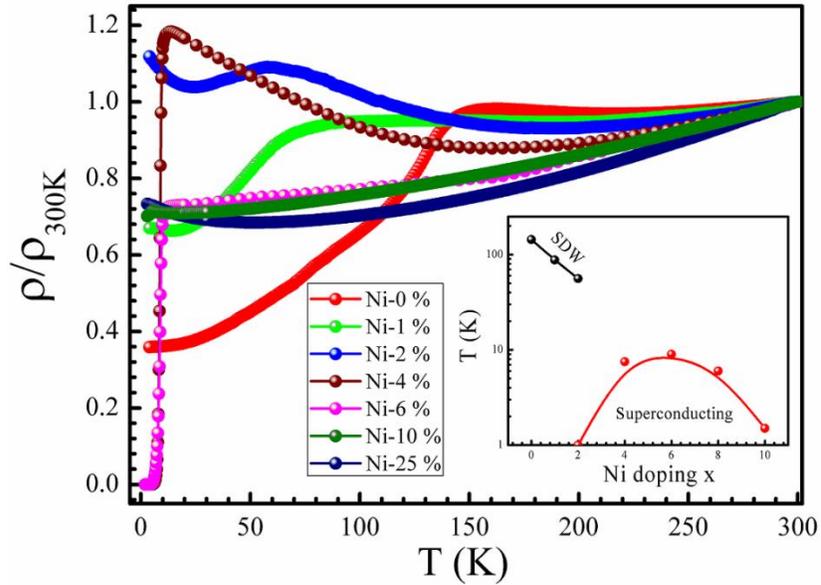

**Figure 2(a):** Resistivity behavior with temperature variation ρ(T) of representative samples of SmFe$_{1-x}$Ni$_x$AsO for $x$= 0.0, 0.01, 0.02,0.04, 0.06, 0.10 and 0.25 at zero field. Inset shows the electronic phase diagram of SmFe$_{1-x}$Ni$_x$AsO.

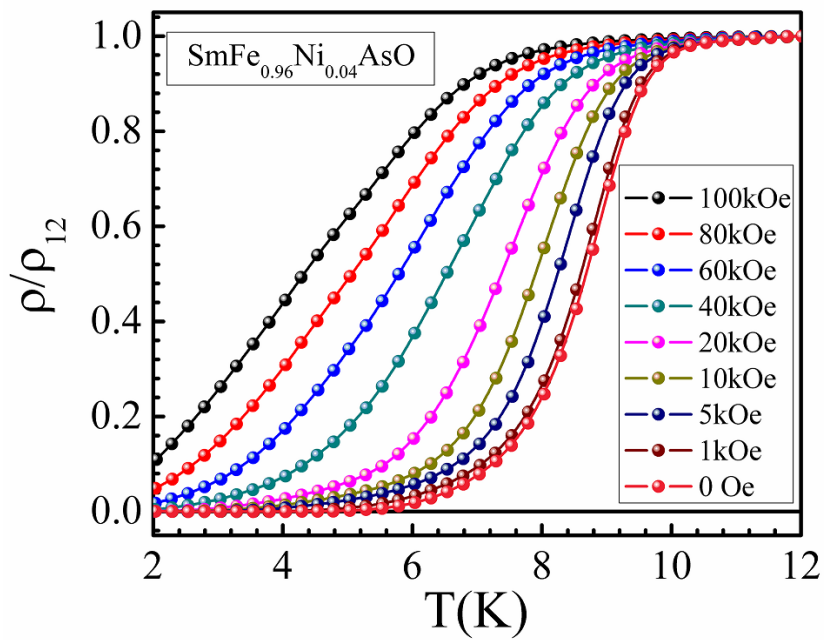

**Figure 2(b):** Resistivity behaviour in the presence of applied magnetic field ρ(T)H up to 10 Tesla for SmFe$_{0.96}$Ni$_{.04}$AsO.

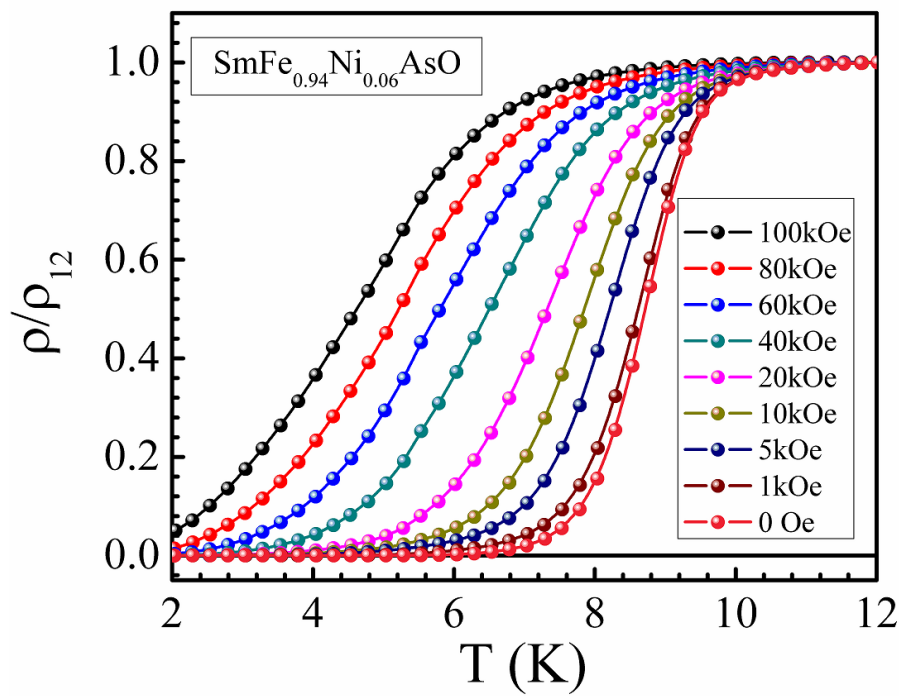

**Figure 2(c):** Resistivity behaviour in the presence of applied magnetic field ρ(T)H up to 10 Tesla for SmFe$_{0.94}$Ni$_{.06}$AsO.

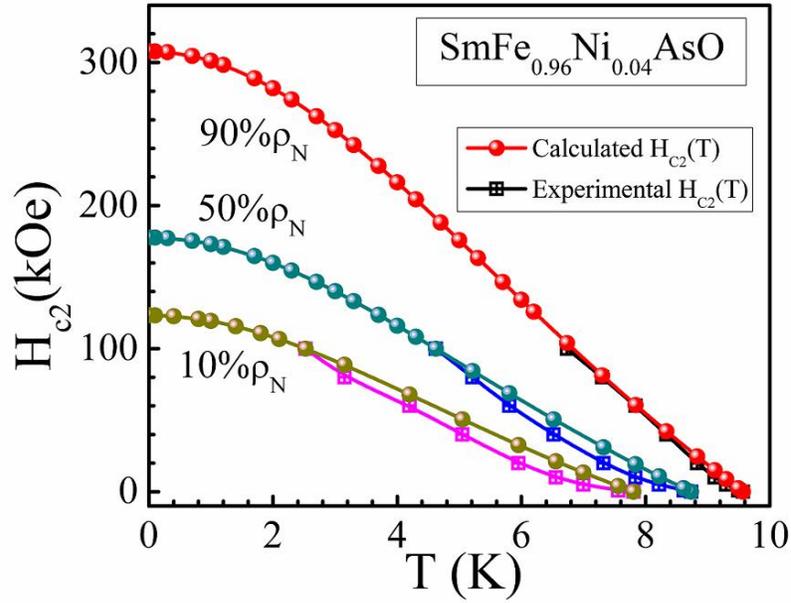

**Figure 2(d):** The upper critical field for the sample SmFe$_{0.96}$Ni$_{.04}$AsO using Ginzburg-Landau (GL) equation for 90%, 50% and 10 % drop of resistance of the normal state resistance.

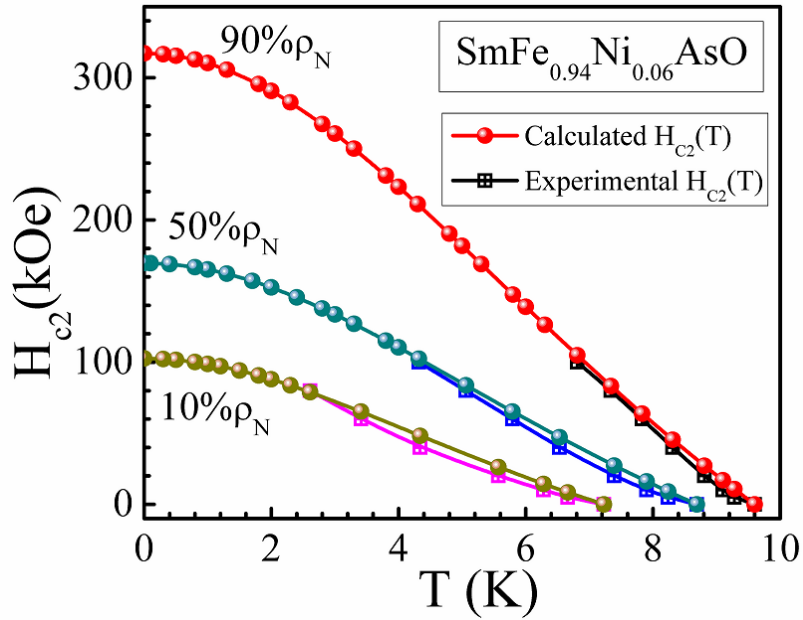

**Figure 2(d):** The upper critical field for the sample SmFe$_{0.94}$Ni$_{.06}$AsO using Ginzburg-Landau (GL) equation for 90%, 50% and 10 % drop of resistance of the normal state resistance.

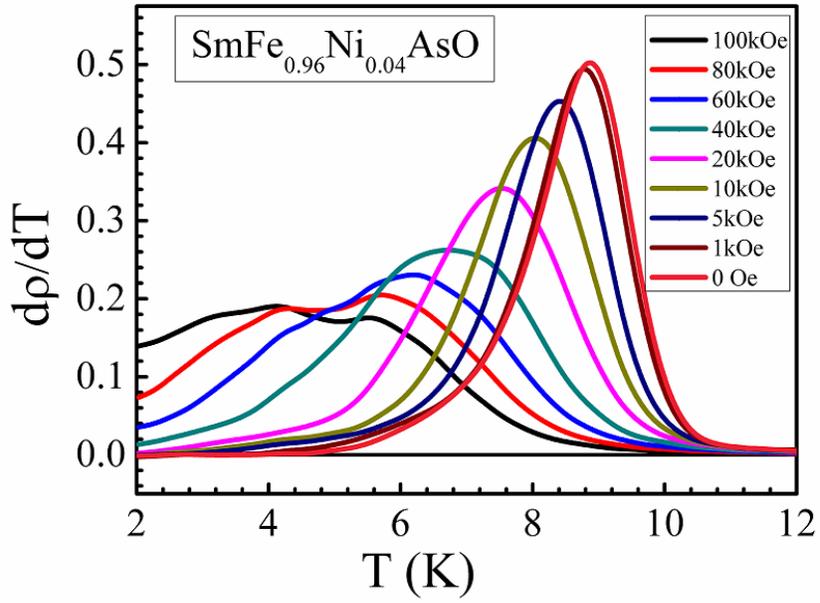

**Figure 2(f):** Temperature derivative of normalized resistivity of SmFe$_{0.96}$Ni$_{.04}$AsO sample.

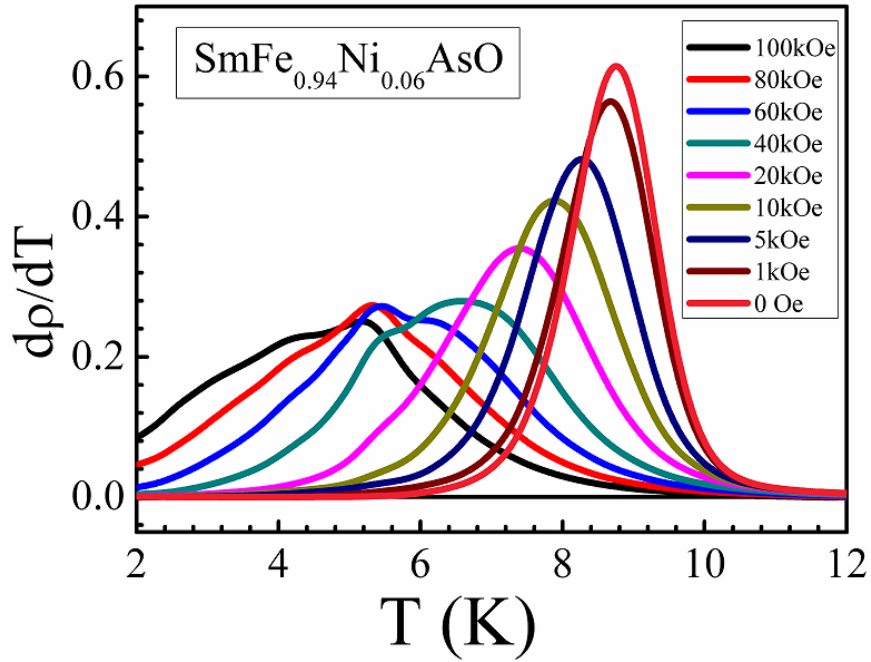

**Figure 2 (g):** Temperature derivative of normalized resistivity of SmFe$_{0.94}$Ni$_{.06}$AsO sample.

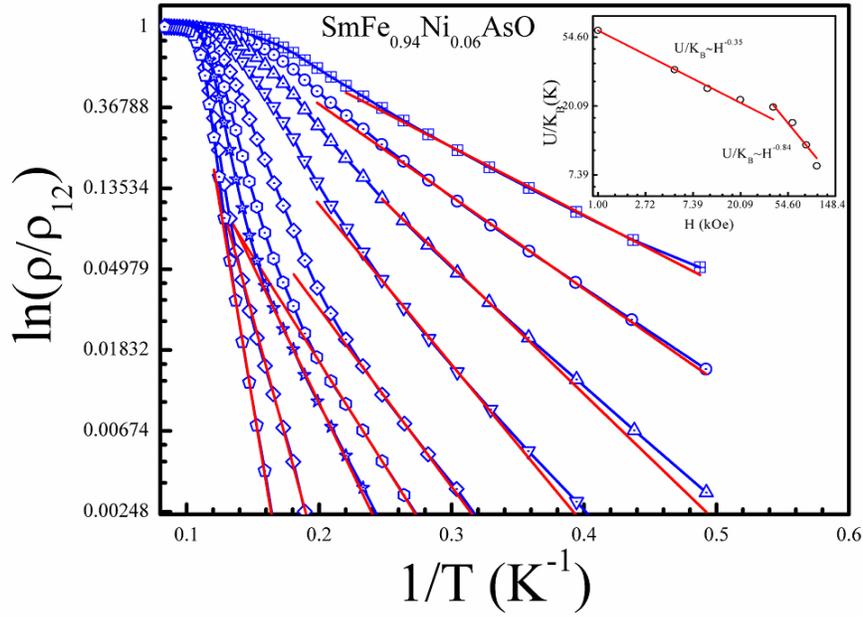

**Figure 3:** Fitted Arrhenius plot of resistivity for SmFe$_{0.94}$Ni$_{.06}$AsO sample. U$_0$ dependence of magnetic field is shown in inset.

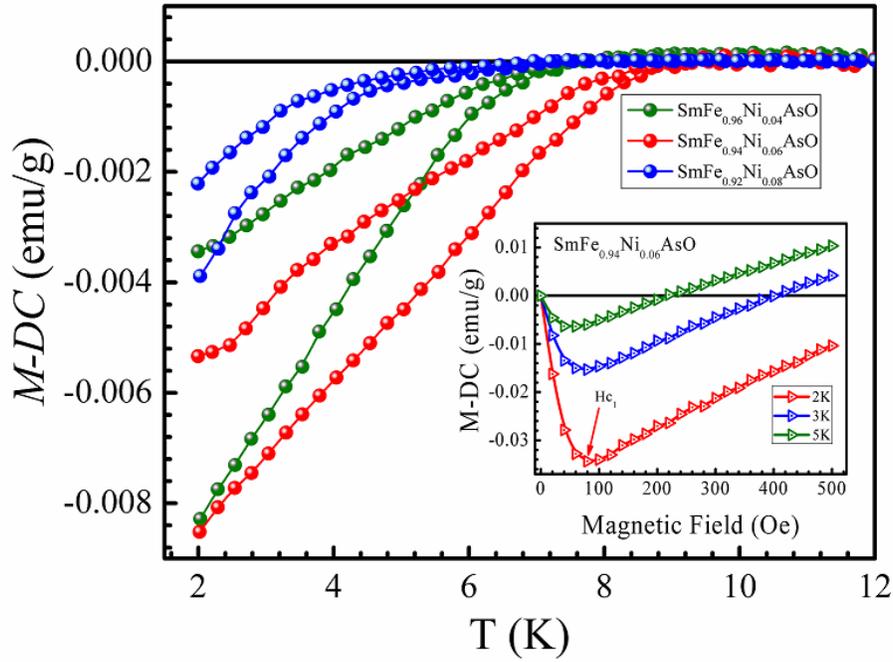

**Figure 4(a):** Temperature variation of magnetic susceptibility M(T) in *FC* and *ZFC* condition for SmFe$_{1-x}$Ni$_x$AsO; x=0.0.04, 0.06 & 0.08 compounds. Inset shows the first quadrant of M(H) loop at 2, 3 & 5 K for same sample.

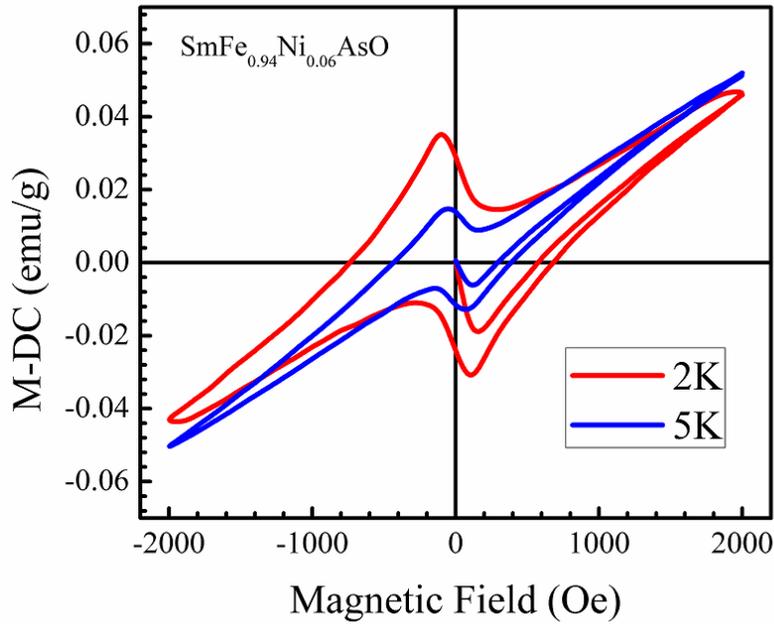

**Figure 4(b):** Variation of Magnetization under applied magnetic field, M (*H*) at 2, & 5 K for the highest T$_c$ sample SmFe$_{0.94}$Co$_{0.06}$AsO

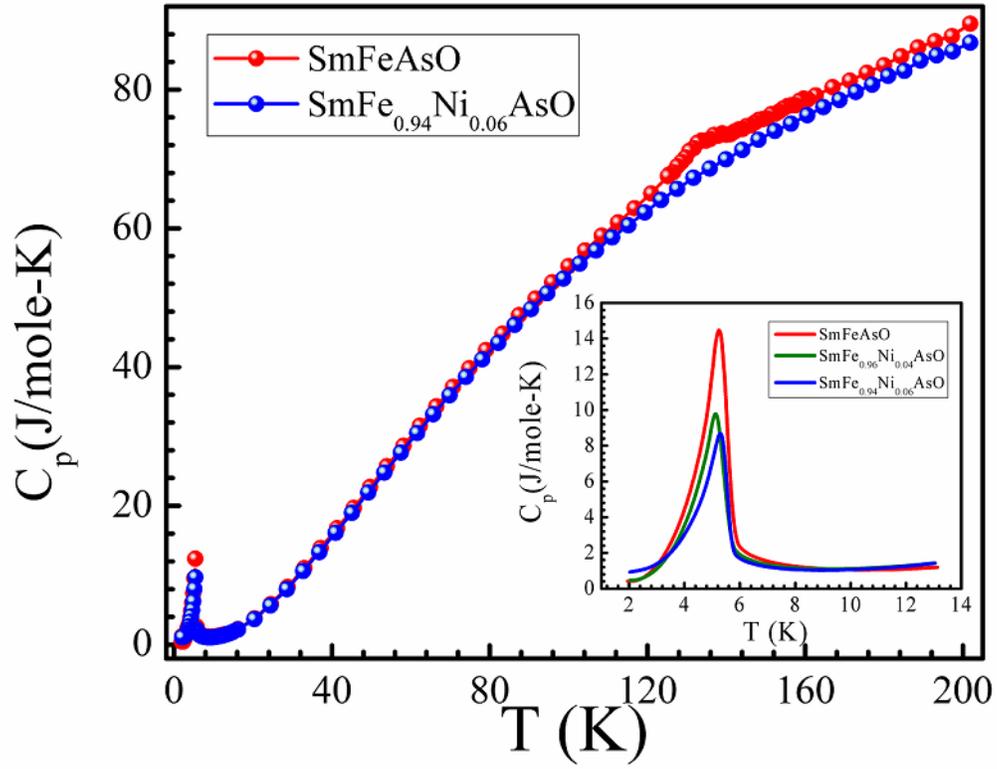

**Figure 5:** Zero field Heat capacity ($C_p$) versus temperature for the SmFeAsO and SmFe$_{0.94}$Co$_{0.06}$AsO sample, and inset shows the suppression of entropy for T$_N$ of Sm with Ni doping